\documentclass[lettersize,journal]{IEEEtran}
\usepackage{amsthm,amsmath,amssymb,amsfonts}
\usepackage{algorithmic}
\usepackage{algorithm}

\usepackage{threeparttable} 
\usepackage{pifont}
\usepackage{booktabs,makecell, multirow, tabularx}
\usepackage{hyperref}
\usepackage{enumitem}
\usepackage{tikz}
\usetikzlibrary{arrows,positioning}
\usepackage{subfigure}
\usepackage{listings}
\usepackage{xcolor}

\newtheorem{theorem}{Theorem}

\newtheorem{definition}[theorem]{Definition}

\newcommand{\Com}[1]{\textcolor{blue!70!black}{{#1}}}

\IEEEoverridecommandlockouts

\begin{document}
\title{VeriFuzzy: A Dynamic Verifiable Fuzzy Search Service Framework for Encrypted Cloud Data}

\author{
Jie Zhang,~\IEEEmembership{Student Member,~IEEE},
Xiaohong Li,~\IEEEmembership{Member,~IEEE},
Man Zheng,
Ruitao Feng,\\
Shanshan Xu, 
Zhe Hou and Guangdong Bai,~\IEEEmembership{Member,~IEEE}

\thanks{This work is supported in part by the National Key Research and Development Program of China under Grant 2023YFB3107103, 2021YFF1201102, in part by the National Natural Science Foundation of China under Grant 62262073, 62332005.}
\thanks{ Jie Zhang, Xiaohong Li and Man Zheng are with the College of Intelligence and Computing, Tianjin University, Tianjin, China. (email: \{jackzhang, xiaohongli, manzheng\}@tju.edu.cn).}
\thanks{ Ruitao Feng is with the Faculty of Science and Engineering, Southern Cross University, Australia (e-mail: ruitao.feng@scu.edu.au).}
\thanks{ Shanshan Xu is with the School of Geographic Sciences, East China Normal University, Shanghai, China. (email: s.xu.ecnu@gmail.com). }
\thanks{ Zhe Hou is with the School of Information and Communication Technology, Griffith University, Nathan, Australia. (email: z.hou@griffith.edu.au).}
\thanks{ Guangdong Bai is with the Department of Computer Science, City University of Hong Kong, Hong Kong, China. (e-mail: baiguangdong@gmail.com).}
\thanks{ Jie Zhang and Man Zheng contributed equally to this work.}
\thanks{  Ruitao Feng and Guangdong Bai are the corresponding authors.}
}

\markboth{Journal of \LaTeX\ Class Files,~Vol.~14, No.~8, August~2021}%
{Zhang \MakeLowercase{\textit{et al.}}: A Sample Article Using IEEEtran.cls for IEEE Journals}

\IEEEpubidadjcol
\maketitle

\begin{abstract}
Enabling search over encrypted cloud data is essential for privacy-preserving data outsourcing. While searchable encryption has evolved to support individual requirements like fuzzy matching, dynamic updates, and result verification, designing a service that supports dynamic, verifiable fuzzy search (DVFS) over encrypted cloud data remains a fundamental challenge due to inherent conflicts between underlying technologies. Existing approaches struggle with simultaneously achieving efficiency, functionality, and security, often forcing impractical trade-offs.


This paper presents \textbf{VeriFuzzy}, a novel DVFS service framework that cohesively integrates three innovations: an \textit{Enhanced Virtual Binary Tree (EVBTree)} that decouples fuzzy semantics from index logic to support $O(\log n)$ search/updates; a \textit{blockchain-reconstructed verification} mechanism that ensures result integrity with logarithmic complexity; and a \textit{dual-repository state management} scheme that achieves IND-CKA2 security by neutralizing branch leakage. 
 Extensive evaluation on 3,500+ documents shows VeriFuzzy achieves 41\% faster search, $5\times$ more efficient verification, and constant-time index updates compared to state-of-the-art alternatives. Our code and dataset are now open source, hoping to inspire future DVFS research.

\end{abstract}

\begin{IEEEkeywords}
Searchable Encryption, Fuzzy Multi-keyword Search, Dynamic Updates, Result Verifiability, Cloud Security, Blockchain, IND-CKA2, Search-As-a-Service.
\end{IEEEkeywords}

\textbf{
\textcolor{blue!70!black}{
This work has been submitted to the IEEE for possible publication. Copyright may be transferred without notice, after which this version may no longer be accessible.
}}

\section{Introduction}
\label{sec:introduction}
The global digital transformation is driving the widespread outsourcing of data to cloud platforms~\cite{flexera2023cloud}. While cloud data services reduce storage costs and improve scalability~\cite{r1}, their inherent characteristics such as openness, centralization, and lack of regulation have raised serious concerns about cloud data security, especially in sensitive areas such as healthcare, finance, and government affairs~\cite{yan2023dynamic,zhang2023backward,Zhang2018Searchable}.
Traditional encryption, though effective in preventing data breaches~\cite{thales2023datathreat}, significantly impairs the ability to perform efficient keyword-based search, which is essential for ensuring timely data access and maintaining quality of service.

While Searchable Encryption (SE) enables searching over encrypted data without revealing its contents~\cite{song2000practical}, real-world applications demand three critical capabilities simultaneously: (i) \textit{fuzzy multi-keyword search} to tolerate typos and variants~\cite{Li2010Fuzzy}, (ii) \textit{dynamic updates} to support real-time data operations~\cite{kamara2013parallel}, and (iii) \textit{result verifiability} to ensure integrity in untrusted environments~\cite{Shao2022Achieve}. 

Three existing mature technologies in SE development can address the above core requirements respectively:
(i) \textit{Fuzzy search} via Locality-Sensitive Hashing (LSH) \cite{FU2016Toward} provides efficient approximate matching by mapping similar keywords to identical buckets with tunable precision; 
(ii) \textit{Dynamic updates} demand tree-based structures like Virtual Binary Trees (VBTree) \cite{wu2019vbtree} to maintain logarithmic complexity for real-time operations; 
(iii) \textit{Result verification} leverages blockchain's immutable ledger and Merkle trees \cite{zhang2021blockchain} for trustless integrity validation. 
Each technique represents the state-of-the-art in its respective domain, making them natural candidates for integration.

However, despite significant progress in individual components of searchable encryption, the Dynamic Verifiable Fuzzy Search (DVFS) schemes that support fuzzy search, dynamic updates, and result verifiability remain challenging to implement, with the core issue stemming from \textit{fundamental incompatibilities} between the above techniques.

\textbf{Challenges of Prior Art in DVFS.} Current solutions face the following core challenges:


\begin{enumerate}[leftmargin=*, noitemsep]
\item \textbf{LSH leads to Semantic-Index Coupling.} 
While fuzzy techniques like locality-sensitive hashing (LSH)~\cite{FU2016Toward} or MinHash~\cite{XIE2024287} can effectively map semantically similar keywords to identical buckets, 
fundamental incompatibility arises when directly applying LSH to dynamic indexing schemes requiring tree structures (e.g., Virtual Binary Tree~\cite{wu2019vbtree}) for $O(\log n)$ operations, causing \textbf{\textit{semantic-index coupling}}, where fuzzy keywords are mapped to fixed positions in tree-based indices, causing any update to document-keyword relationships to trigger cascading index reorganizations with $O(n^2)$ complexity \cite{Zhong2020Efficient}.
This architectural conflict forces existing schemes to choose between fuzzy search capability and practical update performance, either sacrifice scalability~\cite{wu2019vbtree,Zhong2020Efficient,li2017adaptively,Liutowards2025,liu2020multi,xuforward2025} or revert to static indexing models~\cite{Shao2022Achieve,FU2016Toward,xieefficient2025,tong2023verifiable}.

\item \textbf{Blockchain has Verification Overhead.} 
Although blockchain systems based on Merkle trees~\cite{li2021blockchain, tong2023verifiable,tong2022vpsl,cuidynamic2025} theoretically provide trustless result verification with $O(\log n)$  complexity in a decentralized environment, such schemes~\cite{zhang2021blockchain,liu2020multi,xu2019vchain,YAN2022103353,WANG2023110045} lead to \textbf{\textit{verification overhead}} imbalance in dynamic environments with frequent index changes. Specifically, additional $O(n)$ complexity is required for tree structure reconstruction during updates, resulting in actual verification overhead growing linearly. This creates an impossible choice: either accept inefficient verification~\cite{cuidynamic2025} or abandon completeness guarantees.

\item \textbf{VBTree Amplifies Branch Leakage.} 
The logarithmic search efficiency of tree structures \cite{wu2019vbtree} comes at the cost of transparent traversal paths, where each query reveals the specific nodes accessed, enabling adversaries to reconstruct query semantics through statistical analysis \cite{bost2016ovarphiovarsigma}. 
Although the proposed Virtual Binary Tree (VBTree) addresses this issue by implementing branch hiding for exact queries through its virtualized structure, it actually exacerbates the \textbf{\textit{branch leakage}} under fuzzy search. This is because VBTree was designed with only individual branch hiding in mind, and cannot address the branch leakage introduced by fuzzy search semantics. 
Attackers can perform path inference through multiple sets of keyword variants to obtain core keywords, and further deduce keyword distribution, relationships, and even specific query identities~\cite{li2017adaptively,tong2023verifiable}. Consequently, most practical schemes either accept this leakage~\cite{wu2019vbtree,Zhong2020Efficient,liu2020multi,xuforward2025} or sacrifice tree-based structures to achieve the current strongest security level (\textbf{IND-CKA2}) \cite{tong2023verifiable}. This creates a security-functionality trade-off where stronger privacy guarantees necessitate sacrificing either search efficiency or query expressiveness.

\end{enumerate}


\textbf{Our VeriFuzzy Framework: A Cohesive Solution.} 
To systematically address these interconnected challenges, we propose VeriFuzzy, a novel dynamic verifiable fuzzy search service framework that transforms rather than merely combines existing components through three foundational innovations:

\begin{itemize}[leftmargin=*, noitemsep]
\item \textbf{Resolving Semantic-Index Coupling:} We resolve the semantic-index coupling through a novel \textit{Enhanced Virtual Binary Tree with Fuzzy Embedding (EVBTree)}. Unlike previous approaches that bind LSH groups to specific nodes, our method treats LSH-generated trapdoors as cryptographic inputs to the VBTree. Our EVBTree index (Algorithm~\ref{alg:index_gen}) stores each LSH bucket independently while maintaining VBTree's logical structure. In the EVBTree, each keyword $w$ is converted to uni-gram vector $\vec{v}$ and processed through $k$ LSH functions, and for each bucket $b \in S_w$, we compute index entries for all nodes on the path to the document's leaf. This architectural separation ensures that: (i) the tree maintains its $O(\log n)$ search/update efficiency independent of semantic groupings, and (ii) LSH buckets are stored as independent entries for fuzzy update without structural modifications. 
This approach maintains $O(\log n)$ search complexity while supporting fuzzy matching through multiple bucket evaluation.




\item \textbf{Bridging the Dynamic-Verification Gap:} Our \textit{blockchain-reconstructed verification} mechanism leverages smart contracts to generate Merkle proofs on-demand during query execution, eliminating the need for complete tree reconstruction during updates. Only cryptographic digests and deletion logs are stored on-chain. Smart contracts reconstruct search paths during verification using Merkle proofs, and historical version management through dual repositories enables forward privacy without verification overhead. We reduce verification complexity to $O(\log n)$ by processing only relevant paths rather than the entire index, while maintaining full completeness guarantees through distributed trust.

\item \textbf{Neutralizing Branch Leakage:} Through \textit{dual-repository state management} and cryptographic chaining of keyword versions, we achieve IND-CKA2 security without structural pattern leakage, enabling efficient updates while preventing adversaries from correlating queries through access path analysis. In search phase (Algorithm~\ref{alg:search}), we first check all historical trapdoor versions to prevent correlation attacks, then terminate branches at first mismatch to reduce observable pattern density, and finally delete the filter result after search completion to minimize pattern leakage. These techniques preserve VBTree's privacy benefits while accommodating fuzzy search performance requirements.
\end{itemize}

Unlike previous integration attempts that suffered from the fundamental conflicts outlined above, VeriFuzzy represents an architectural breakthrough that rethinks the relationship between functionality, efficiency, and security in encrypted search systems.
Our implementation on a corpus of 3,500+ documents demonstrates: (i) 41\% faster search than STATE-of-the-art fuzzy SSE schemes, (ii) $5\times$ verification efficiency compared to linear-verifiable models, and (iii) full support for dynamic updates with constant-time index refresh.

\textbf{Summary of Contributions:}
\begin{itemize}
    \item We design a dynamic encrypted search framework supporting efficient fuzzy multi-keyword queries and real-time updates via a novel virtual binary tree structure.
    \item We develop a blockchain-based verification mechanism that ensures completeness and correctness with $O(\log n)$ complexity.
    \item We formally prove the security of VeriFuzzy and demonstrate its practicality through comprehensive experimental evaluation. 
    \item While our open-source release does not expose proprietary blockchain chaincode (due to commercial deployment restrictions), our code and dataset are publicly available at: \url{https://github.com/JackAugust/VeriFuzzy.git}, which are used for implementing dynamic fuzzy search and other functionalities in VeriFuzzy.
\end{itemize}

The remainder of this paper is organized as follows: Section~\ref{sec:related-work} reviews related work; Section~\ref{sec:preliminaries} presents preliminaries; Section~\ref{sec:system-model} describes our system model; Section~\ref{sec:scheme} details the VeriFuzzy scheme; Section~\ref{SecurityAnalysis} provides security analysis; Section~\ref{sec:performance} presents experimental results; Section~\ref{discussion} provides discussions and future work prospects, and Section~\ref{sec:conclusion} concludes.

\section{Related Work}

\subsection{Fuzzy Searchable Encryption}
The limitation of precise keyword matching in addressing spelling errors in query keywords, which leads to incorrect outputs, has created an urgent need for designing keyword search schemes that support approximate matching, thus introducing the concept of fuzzy searchable encryption. Li et al. \cite{Li2010Fuzzy} proposed the first fuzzy keyword search scheme that utilizes edit distance and wildcards to construct a predefined fuzzy set to cover spelling errors; however, it only supports single-keyword queries. Wang et al. \cite{wang2014privacy} employed Bloom filters and LSH to facilitate fuzzy matching searches. Building on this,
Fu et al. \cite{FU2016Toward} improved \cite{wang2014privacy} scheme by introducing the uni-gram model to improve the accuracy of fuzzy ranked search. 
Liu et al. \cite{liu2022prime} implemented an effective wildcard-based fuzzy multi-keyword search using edit distance that supports both AND and OR search semantics. 
Xie et al. \cite{xieefficient2025} pioneered geohash-primes indexing for spatial fuzzy queries. Liu et al. \cite{Liutowards2025} achieved $O(k)$ conjunctive search via binary tree indexing. 
However, all of the above schemes only support static data and ignore the dynamic scenario that does not meet the actual production requirements.

\subsection{Dynamic SSE}
The demand for real-time data processing has motivated Dynamic SSE (DSSE) research.
Kamara et al. \cite{kamara2013parallel} designed a keyword red-black tree index, realizing a sublinear DSSE scheme that supports efficient updates. Xia et al. \cite{xia2016secure} constructed a special keyword balanced binary tree as the index that supports flexible dynamic operation. 
Li et al. \cite{li2017adaptively} introduced an indistinguishable binary tree and an indistinguishable bloom filter data structure for conjunctive query.
Wu et al. \cite{wu2019vbtree} further proposed the Virtual Binary Tree (VBTree) structure, achieving scalable index sizes and sublinear update times. Zhong et al. \cite{Zhong2020Efficient} utilized LSH and balanced binary trees (which assemble all Bloom index vectors) to achieve fault-tolerant multi-keyword fuzzy retrieval and support dynamic file updates, Xu et al. \cite{xuforward2025} enabled multi-user verification, yet integrating these two functionalities may impact search efficiency. 

\subsection{Verifiable SSE}
A significant body of research has addressed data correctness and integrity in response to malicious cloud services. Liu et al. \cite{liu2022enabling} employed RSA accumulators to generate proofs for search results, although this approach incurs substantial computational overhead. Wan et al. \cite{wan2018vpsearch} utilized homomorphic MAC techniques to authenticate each index, thereby verifying the correctness of relevance score calculations. Tong et al. \cite{tong2022vpsl} proposed an authenticated tree-based index structure based on merkle hash trees. Zhang et al. \cite{zhang2021blockchain} leveraged blockchain technology to achieve certificate-free public verification of data integrity without a central authority. Regarding correctness and integrity verification, Liu et al. \cite{liu2020multi} employed RSA accumulators to authenticate index structures; however, this method is not suitable for tree-based index structures. Shao et al. \cite{Shao2022Achieve} implemented efficient verifiable fuzzy multi-keyword search using keyed hash message authentication codes and cardinality trees. Li et al. \cite{li2021blockchain} and Cui et al. \cite{cuidynamic2025} introduced chain-indexed Merkle trees for verifiable fuzzy search. Tong et al. \cite{tong2023verifiable} advanced fuzzy verifiability.  

\subsection{Integrated Solutions and Research Gaps}
As synthesized in Table~\ref{tab:comprehensive-comp}, our VeriFuzzy represents the first solution that simultaneously achieves all three objectives without compromising security or efficiency.
Compared with Tong \textit{et al.}'s VFSA\cite{tong2023verifiable}, we achieve dynamic updates and reduce verification complexity from $O(n)$ to $O(\log n)$ without compromising security.
Compared to Zhong \textit{et al.}'s EDMF\cite{Zhong2020Efficient}, our scheme adds verifiability and eliminates the $O(n^2)$ update bottleneck.
Cui \textit{et al.}'s DVFKF \cite{cuidynamic2025} represents the closest prior art. However, its $O(n)$ search complexity and lack of completeness verification leave critical gaps. Our work  finally achieves the quadripartite goal of \textbf{fuzzy search}, \textbf{dynamic updates}, \textbf{dual verification}, and \textbf{sublinear complexity} simultaneously.

\label{sec:related-work}
\begin{table}[t]
\caption{Comparison of encrypted search schemes. $F_1$: Fuzzy multi-keyword; $F_2$: Dynamic updates; $F_3$: Correctness verification; $F_4$: Completeness verification; $F_5$: Search complexity; $F_6$: Verification complexity; $F_7$: Security model.}
\label{tab:comprehensive-comp}
    \centering
    \resizebox{0.5\textwidth}{!}{
    \begin{tabular}{ccccccccc}
    \hline
\textbf{Scheme}& $F_1$ & $F_2$ & $F_3$  &$F_4$ 
& $F_5$ &$F_6$ &$F_7$ \\ \hline 
\cite{FU2016Toward}&       $\checkmark $&       $\times$&       $\times$ &$\times$
&       $O(n)$ &- &$<$KPA \\
 \cite{wu2019vbtree}& $\times$& $\checkmark $& $\times$ & $\times$
&$O(\log n)$ &-  & IND-CKA2 \\
 \cite{li2017adaptively} & $\times $& $\checkmark $& $\times$ & $\times$ &$O(n\log n)$ &- &IND-CKA2 \\ 
 \cite{Zhong2020Efficient}& $\checkmark $ & $\checkmark $ & $\times $ & $\times$ &$O(\log n)$ &- & KPA \\
 \cite{xieefficient2025} & $\checkmark $ & $\times $ & $\times $ & $\times $ &  $O(\log n)$  & - &IND-CKA \\
\cite{Liutowards2025} & $\times $ & $\checkmark $ & $\times $ & $\times $ & $O(\log n)$ & - &	IND-CKA2 \\
\cite{liu2020multi}&       $\times$&       $\checkmark $&       $\checkmark $ &$\checkmark $ &       $O(\log n)$ &$O(n)$ & UC-CKA \\
 \cite{xuforward2025}& $\times $& $\checkmark$& $\checkmark $ & $\checkmark $ & $O(n)$ &$O(n)$ &IND-CKA \\
\cite{tong2023verifiable} & $\checkmark $ & $\times $ & $\checkmark $ & $\checkmark $ & $O(\log n)$ & $O(n)$ &	IND-CKA2 \\
\cite{cuidynamic2025} & $\checkmark $ & $\checkmark $ & $\checkmark $ & $\checkmark $&  $O(n)$ & $O(n)$&IND-CKA \\
Ours &       $\checkmark $&       $\checkmark $&       $\checkmark $ &$\checkmark $&       $O(\log n)$ &$O(\log n)$ &	IND-CKA2\\ \hline

    \end{tabular}
}
\end{table}

\section{Preliminaries}
\label{sec:preliminaries} 
In this section, we review the relevant background knowledge, including locality sensitive hashing \cite{wang2014privacy},virtual binary tree \cite{wu2019vbtree} and blockchain \cite{nakamoto2008bitcoin}.
\subsection{Locality Sensitive Hashing}\label{AA}
\textbf{Locality Sensitive Hashing (LSH)} LSH solves approximate or exact Near Neighbor Search in high dimensional space. It hashes inputs, likely mapping similar items to same buckets. Here, keywords are LSH-processed and then added to indexes or tokens for fuzzy search. A hash function family $H$ is ($R_1$,$R_2$,$P_1$,$P_2$ )-sensitive when for any $p$, $q$ and $h \in H$: 

\begin{itemize}
\item 	if $d(p,q) \leq R_1$,  $Pr[h(p)=h(q)] \geq P_1$;
\item 	if $d(p,q) \geq R_2$,  $Pr[h(p)=h(q)] \leq P_2$;
\end{itemize}

Here, $d(p,q)$ is the distance between $p$ and $q$, $R_1,R_2$ are the distance thresholds ($R_1 < R_2$), and $P_1,P_2$ are the prob thresholds ($P_1 > P_2$). In this work, for similar keyword identification, we use $p$-stable LSH \cite{datar2004locality} mapping $d$ dimensional vectors to integers. The $p$-stable function in our scheme is $h_{(\mathbf{a},b)} (\vec{v} )=\lfloor{\frac{\mathbf{a} \cdot \vec{v}+b} {c}}\rfloor$. Here, $\mathbf{a}$ is a $d$ dimensional vector with each dim randomly and independently chosen from the $p$-stable dist, $b\in[0,c]$ is random, and $c$ is a fixed constant per family. Varying $\mathbf{a}$ and $b$ generates different $p$ stable LSH functions.

\subsection{VBTree}\label{AA}
\textbf{Virtual Binary Tree (VBTree) }VBTree, first introduced in \cite{wu2019vbtree}, is a full binary tree that exists logically rather than physically. Its key design is top-down indexing element organization without storing branches or nodes. Each VBTree node has a path: for non-terminal nodes, a '0' is appended to the left child's path and a '1' to the right child's path. For node $v$, $path(v)$ is the binary string from root to $v$ (root's $path(v)$ is empty). Let $leaf_i$ be the $i^{th}$ leaf and $Nodes(i)$ be the nodes from the root to $leaf_i$. To index keyword $w$ for doc $i$, insert $H(path(v)\|F_K(w))$ into each $Nodes(i)$ node, where $H$ is a random oracle and $F_K$ is a pseudo-random function. 

\begin{figure}
\centering
\includegraphics[width=3in]{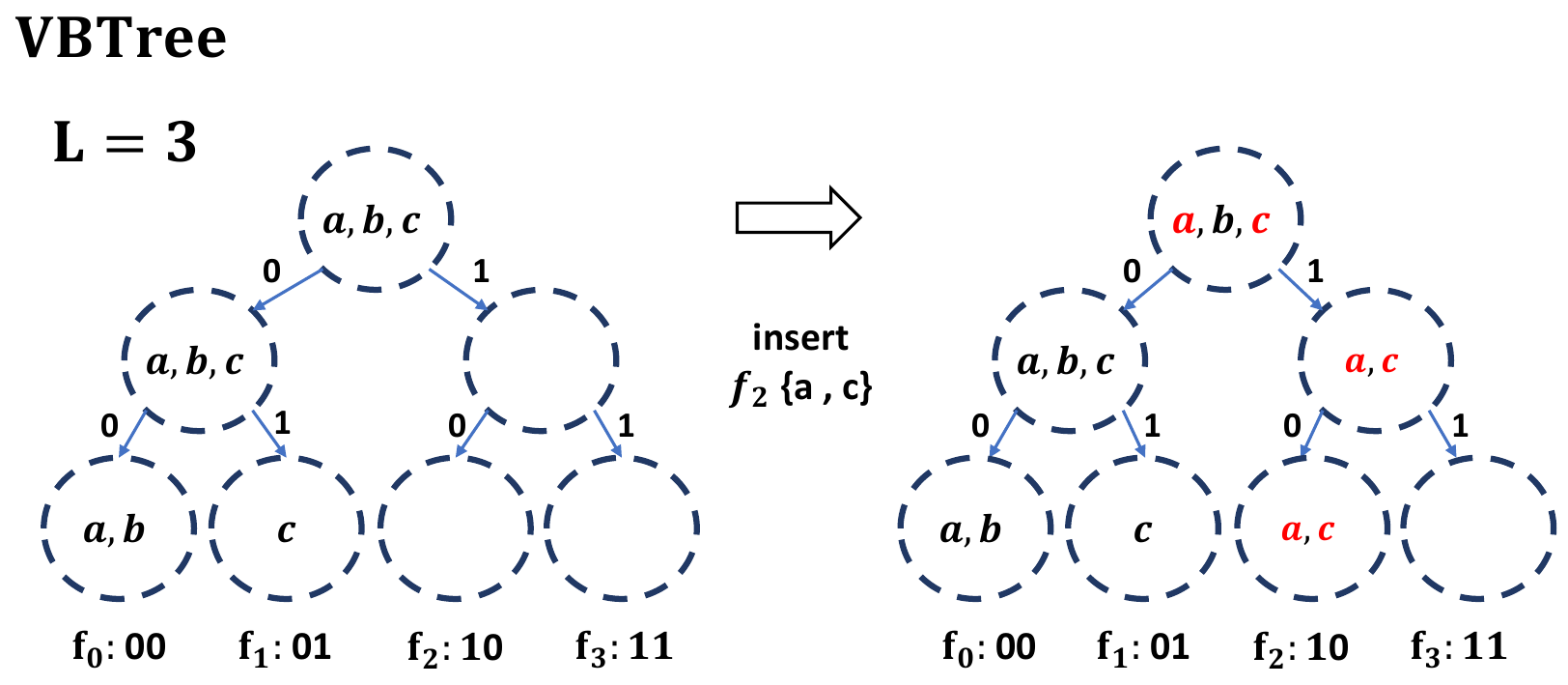}
\caption{An example of vbtree.}
\label{vbtree}
\end{figure}

Fig.~\ref{vbtree} shows an example of a virtual binary tree of height $L = 3$. There are four leaves in the tree. $path(leaf_1 )$ denotes string ``01''. $Nodes(1)$ denotes a set of tree nodes $\{root,node_0,node_{01}\}$. The values of keyword ``c'' of file identifier ``1'' in the hash table are a set of key-value pairs, i.e., denoted $W($``c''$,1)=\{H(path(v) \| F_K (c))\}_{v \in Nodes(1)}=\{H($``''$\| F_K (w)),H($``0''$ \| F_K (w)),H($``01''$ \| F_K (w))\}$. All key-value pairs in this VBTree are $W($``a''$,0) \cup W($``b''$,0) \cup W($``a''$,1) \cup W($``b''$,1) \cup W($``c''$,1)$. To insert a file $f_2$ with two keywords $\{a, c\}$, the data owner converts $f_2$ to the path ``10'' and inserts the encrypted items from the root to the leaf along path ``10''. The inserted items are $W($``a''$,2) \cup W($``c''$,2)$.

For a VBTree storing $n$ real files, with all leaves containing $N$ keywords (where $N$ is the total keyword appearances in all files and $L$ is the tree height), its construction time is $O(LN)$ and index size is $O(\beta N)\approx O(N \log n)$ ($\beta \in[2,L]$). Also, for a $u$-dimensional keyword conjunctive query $q$, the top-down search algorithm has a query complexity of $O(\log_2n)$.

\subsection{Blockchain}\label{AA}
\textbf{Blockchain and Smart Contract}: Blockchain, as a continuously evolving distributed ledger technology, constitutes the fundamental infrastructure of cryptocurrency systems~\cite{nakamoto2008bitcoin}. Its security is underpinned by cryptographic hashing and consensus mechanisms. In a blockchain, each block acts as a permanent record of transactions, with millions of blocks interconnected to form a blockchain network. The consensus mechanism ensures the integrity and consistency of the entire blockchain network, thereby preventing data tampering~\cite{zhang2025isolation}. Originating from Nick Szabo \cite{szabo1996smart}, smart contracts are digital legal agreements executed as computer programs. Intended to foster trust without TTPs, their implementation was hindered by the absence of programmable digital systems. These are unmodifiable scripts on the blockchain. Once deployed in a block, they automatically execute according to predefined logic, independent of a central authority. 

\section{System Model}
\label{sec:system-model}
In this section, we formalize the system components and security requirements for addressing the fundamental challenges identified in Section~\ref{sec:introduction}. Specifically, we aim to resolve the three core conflicts in dynamic verifiable fuzzy search (DVFS): (1) the semantic-index coupling between LSH-based fuzzy search and tree-based dynamic indexing, (2) the verification overhead bottleneck in dynamic environments, and (3) the branch leakage vulnerability amplified by fuzzy query patterns in Virtual Binary Tree (VBTree). 

To systematically address these challenges, this section establishes the foundational framework through four key components: first, we define the system model involving data owners/users, cloud servers, and blockchain entities; second, we formalize the design goals encompassing fuzzy search capability, dynamic updates, verifiable results, and strong security guarantees; third, we provide precise algorithmic definitions for the six core operations of our scheme; finally, we present the security model including adaptive adversary capabilities and IND-CKA2 security notions. Collectively, these components provide the formal underpinnings for our VeriFuzzy framework, enabling the integrated solution described in subsequent sections.

\subsection{The System Model}

As shown in Fig.\ref{fig3}, our system model consists of four types of entities: data owners, data users, cloud servers, and blockchain. The roles of the entities are as follows:

\begin{figure}
\centering
\includegraphics[width=0.5\textwidth]{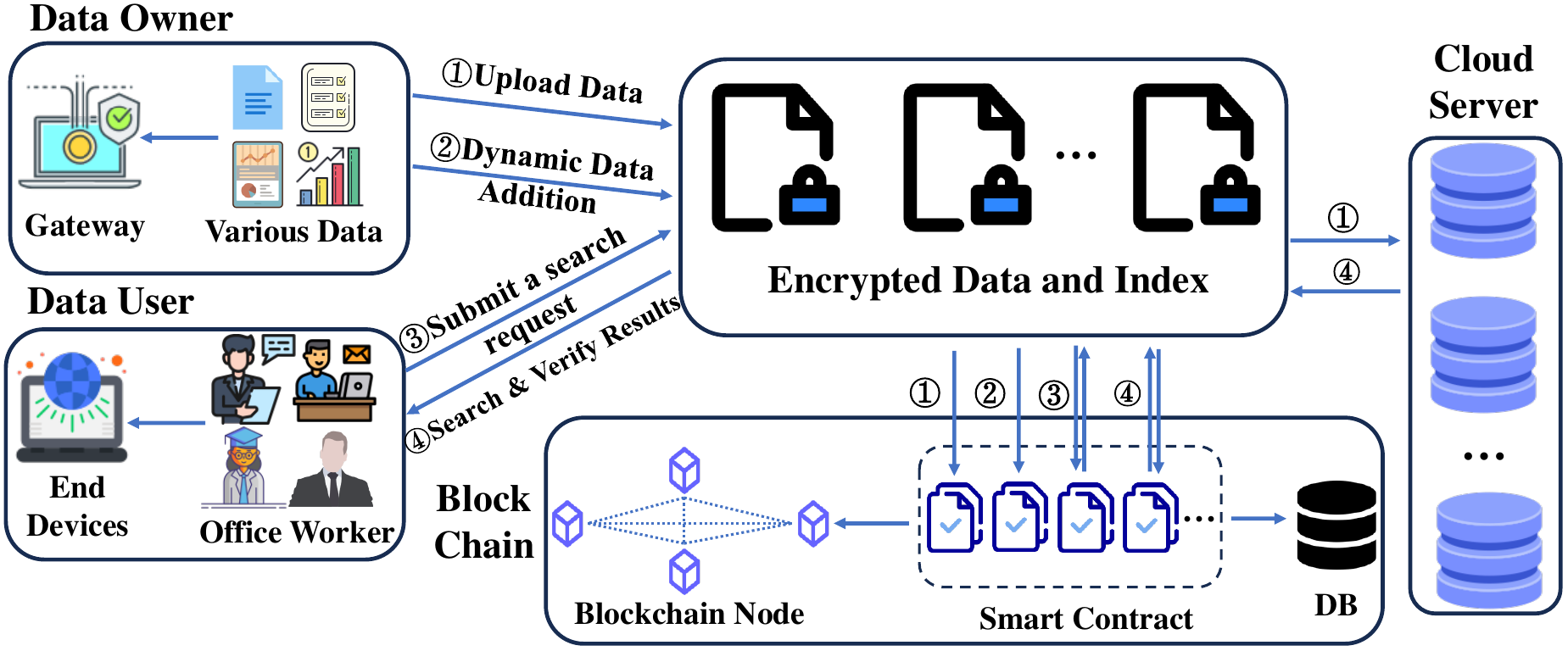}
\caption{System model.}
\label{fig3}
\end{figure}

\begin{itemize}
    \item \textbf{Data Owners (DOs)}: Departments within an enterprise act as data owners, uploading project documents, encrypting them using traditional symmetric encryption algorithms, constructing encrypted indexes, and then uploading them to cloud server and blockchain respectively.
    \item \textbf{Data Users (DUs)}: Employees in each department serve as data users, sending query requests or new document materials to blockchains and cloud server according to their needs.
    \item \textbf{Cloud Server (CS)}:  CS is responsible for storing the uploaded encrypted data and returning the required search results, providing support for cross departmental data interaction.
    \item \textbf{Blockchain (BC)}:  BC stores encrypted indexes and executes query and addition operations on searchable encrypted indexes with the help of smart contracts. During the query process, the search contract records auxiliary information, and the verification contract verifies the correctness and integrity of the results based on this information to ensure data security and consistency.
\end{itemize}

 \subsection{Design Goals}
Our design goals encompass the following key aspects:
\begin{itemize}
  
    \item[\textbf{G1}] \textbf{Robustness.} VeriFuzzy must support both fuzzy multi-keyword search and dynamic data management to enhance robustness in complex business scenarios. Similar to~\cite{Zhong2020Efficient,tong2023verifiable,cuidynamic2025}, it must achieve high accuracy in fuzzy search even when keywords contain common spelling errors (see Section~\ref{RQ1}), and remain effective after updates.
    \item[\textbf{G2}] \textbf{Verifiability.} VeriFuzzy supports the verification of the correctness and completeness of search results, ensuring that the returned data are accurate and reliable.
    \item[\textbf{G3}] \textbf{Efficiency.} VeriFuzzy can achieve sublinear time complexity for search and verification. Compared to traditional solutions, it effectively reduces search response times and enhances verification efficiency, optimizing system performance.
    \item[\textbf{G4}] \textbf{Security and Privacy.}  VeriFuzzy must satisfy adaptive security and ensures forward privacy \cite{li2019searchable} for dynamic updates, ensuring that cloud servers cannot infer any sensitive information from encrypted data documents, indexes, user search tokens, or data updates. 
\end{itemize}

 
\subsection{Algorithm Definition}
The following presents the algorithm definitions of the six algorithms that make up the research scheme of this paper:

$\mathbf{Setup}(1^\lambda,k)\rightarrow Params$: Given a security parameter $\lambda$ and an element $k$, the DO outputs the parameter $Params$.

$\mathbf{IndexGen}(F,C,Params)\rightarrow I$: Given a document set $F$, ciphertext set $C$ and a parameter $Params$, DO outputs an index tree $I$.

$\mathbf{Update}(I,i,W)\rightarrow I'$: Given index construction $I$, a file identifier $ i$, an additional set of keywords $W$, DO outputs the updated index $I'$.

$\mathbf{TrapGen}(Q,Params)\rightarrow TK$: Given a search query Q and parameter $Params$, DU outputs a trapdoor $TK$ for retrieval.

$\mathbf{Search}(I,TK,Params)\rightarrow \{R,AP,D\}$: Given the index tree $I$, the trapdoor $TK$ and the parameter $Params$, BC outputs search results $R$, the auxiliary proof information $AP$ and ciphertext digest information set $D$.

$\mathbf{Verify}(R,AP,D,E,Params)\rightarrow 1/0$: Given the search results $R$, the auxiliary proof information $AP$, the digest information set $D$, the ciphertext set $E$ and the parameter $Params$, BC outputs $1$ if $E$ are correct and complete; otherwise, it returns $0$.

\subsection{Security Definition}

The IND-CKA2 model, first proposed by \cite{r19}, combines indistinguishability and adaptive keyword attacks to ensure stronger confidentiality. Unlike other security models, IND-CKA2 comprehensively evaluates the security of searchable encryption schemes based on leakage functions, ensures that even if an adversary queries a random oracle and selects keywords, other index keywords stay hidden. This model provides a more comprehensive analysis of attackers' behavior and strategies and offers higher security than other security models.

We consider the strongest threat model in searchable encryption: an \textbf{adaptive adversary} $\mathcal{A}$ that operates as follows:

\begin{definition}[Adaptive Adversary]\label{PPT-A-Def}
An adaptive adversary $\mathcal{A}$ is a probabilistic polynomial-time (PPT) algorithm that performs the following:
\begin{enumerate}
\item \textbf{Adaptively chooses queries}: Based on all previously observed information (indexes, trapdoors, search results), $\mathcal{A}$ can strategically select subsequent search queries $Q_i$ and update operations to maximize information leakage.

\item \textbf{Dynamically interacts with the system}: $\mathcal{A}$ engages in multiple rounds of the security game, where each query $Q_i$ may depend on the outcomes of all previous queries $\{Q_1, \dots, Q_{i-1}\}$ and their corresponding trapdoors $\{TK_1, \dots, TK_{i-1}\}$.

\item \textbf{Observes all cryptographic outputs}: $\mathcal{A}$ can observe the encrypted index $I$, all generated trapdoors $TK_i$, search results, and updated indexes $I_{up}$ throughout the entire system lifetime.

\item \textbf{Attempts to distinguish real from simulated worlds}: The adversary's goal is to distinguish between interactions with the real scheme and a simulator that receives only the predefined leakage functions $L = (L_1, L_2, L_3)$.
\end{enumerate}
\end{definition}


Our leakage function $L=(L_1,L_2,L_3)$ follows the standard formulation in most SE schemes \cite{li2017adaptively,Liutowards2025,liu2022enabling,r20}:

\begin{itemize}
\item 	 $L_1 = L_1(F)=(n, sizes, ids, L, M)$. Here, $L_1$ leakage encompasses $M$ (EVBTree hashtable entry count), $L$ (EVBTree height), $n$ (file count), $sizes$ (set of file sizes encrypted by a CPA-secure scheme), and $ids = \{id_1,\cdots,id_n\}$ (set of file identifiers from $1$ to $n$). 
\item $L_2 = L_2(F,Q)=\{SP(w),HIST(w)\}$, which includes the search pattern $SP(w)$ (client's trapdoor repetition to BC) and the history $HIST(w)$ of queried keywords, representing returned encrypted document identities. For a keyword $w$, $SP(w)$ has all search trapdoors for $w$, and $HIST(w)$ is the collection of $w$ containing document identifiers.
\item $L_3 = L_3(W, id)$: This leakage represents the information revealed during encrypted index updates. Here, $W$ and $id$ denote adding keyword set $W$ to the document with identifier $id$. 
\end{itemize}

\begin{definition}[IND-CKA2($L_1,L_2  ,L_3 $)-secure]\label{IND-CKA2-Def}
If for any PPT adaptive adversary $\mathcal{A}$, there exists a simulator $S$ such that $|Pr[Real_A (\lambda)]-Pr[Sim_{A,S} (\lambda)]|\leq negl(\lambda)$, where $negl( )$ denotes a negligible function, then the proposed scheme in this paper is IND-CKA2($L_1,L_2  ,L_3 $)-secure.
The probabilistic experiments $Real_A (\lambda)$ and $Sim_{A,S} (\lambda)$ are shown as follows:

\begin{itemize}
\item 	$Real_A (\lambda)$:  the challenger $C$ runs $\mathbf{Setup}(1^\lambda,k)$ to generate parameter $Params$, A outputs the document set $F=\{f_1  ,\cdots,f_n \}$ and the corresponding keyword set $W=\{ W(f_1 )  ,\cdots ,W(f_n )\}$, and receives the index tree $I$ and encrypted document set $C=\{ c_1  ,\cdots,c_{n} \}$ from the challenger $C$ such that $I\leftarrow\mathbf{IndexGen}(F,C,Params)$. Based on $I$, $\mathcal{A}$ chooses a search query $Q_1$. Taking it as input, the challenger $C$ runs $\mathbf{TrapGen}(Q_1,Params)$ to generate a trapdoor $TK_1$. Based on $I$, $Q_1$, $TK_1$, $A$ chooses a search query $Q_i (2\le i\le s)$, the challenger $C$ repeats the above query process to generate a trapdoor $TK_i (2\le i\le s)$. This process is repeated in a polynomial number of times $s$. $A$ obtains the search trapdoor set $\{TK_1,\cdots,TK_s\}$. Similarly, $A$ sends an updated keyword set $W_j (1\le j\le k)$, and this process is repeated $k$ times. the challenger $C$ runs $\mathbf{Update}(I,i,W_j)$ to generate an updated encrypted index $I_{up}$ after $k$ updates and provides it to the adversary $A$. Finally, $A$ outputs the encrypted index $I$, the updated index $I_{up}$ and the set of search trapdoors $\{TK_1,\cdots,TK_s\}$. 
\item 	$Sim_{A,S} (\lambda)$: $\mathcal{A}$ outputs the document set $F=\{f_1,\cdots,f_n\}$ and the corresponding keyword set $W=\{W(f_1),\cdots,W(f_n)\}$. Given $L_1(F)$, the simulator $S$ generates the ciphertext set $C'=\{c_0',c_1',\cdots,c_{n}'\}$ and the encrypted index $I'$, and sends them to the adversary $\mathcal{A}$. Based on $I'$, $A$ chooses a search query $Q_1$, $S$ is given $L_2 (F,Q_1 )$ and outputs an appropriate trapdoor $TK_1'$. Based on $I'$, $Q_1$, $TK_1'$, $A$ chooses a search query$ Q_i (2\le i\le s)$, $S$ is given $L_2 (F,Q_1,\cdots,Q_i )$ and returns an appropriate trapdoor $TK_i' (2\le i\le s)$ to $A$. This process is repeated $s$ times, and the adversary $A$ obtains the set of search trapdoors $\{TK_1',\cdots,TK_s'\}$. Similarly, the adversary $A$ sends an update keyword set $W_j (1\le j\le k)$, and this process is repeated $k$ times. Based on the given leakage function $L_3 (W_j,id)$, where $id$ is the identifier of the document that the update keyword set $W_j$ belongs to, $S$ sends an updated encrypted index $I_{up}'$ after $k$ updates to $A$. Finally, $A$ outputs the encrypted index $I'$, the updated index $I_{up}'$ and the set of search trapdoors $\{TK_1',\cdots,TK_s'\}$.
\end{itemize}
\end{definition}

This Definition~\ref{IND-CKA2-Def} formalizes the adaptive security of our searchable encryption scheme under the IND-CKA2 model. The security is parameterized by three leakage functions $L_1$, $L_2$, and $L_3$ that quantify the information intentionally revealed to the adversary. The core idea is that for any PPT adversary $A$, the views in the real world (interacting with the actual scheme) and the simulated world (interacting with a simulator that only uses the allowed leakage) are computationally indistinguishable.

\section{Our VeriFuzzy scheme}\label{sec:scheme}

This section details the design of our scheme, closely adhering to the specified objectives. For robustness (\textbf{G1}), VeriFuzzy achieves this through fuzzifying index keywords and enhanced virtual binary tree construction, enabling spelling error tolerance and dynamic data updates; for verifiability (\textbf{G2}), ensured by ciphertext digest recalculation and search tree reconstruction; for efficiency (\textbf{G3}), Implemented via trapdoor computation and top-down search algorithms for optimized query processing; for security and privacy (\textbf{G4}), Security Analysis \ref{SecurityAnalysis} will be fully explained.

\subsection{Setup}\label{5-1}
Given the security parameters $\lambda$ and $k$, the DO outputs the parameter set $Params=\{F_K,H_1,LSH_k\}$, where $F_K$ is a keyed-hash function: $\{0,1\}^*\times\{0,1\}^\lambda \rightarrow\{0,1\}^*$, $H_1$ is a hash function: $\{0,1\}^*\rightarrow\{0,1\}^*$, and $LSH_k$ is a locality-sensitive hash family of $k$ functions $\{h_{a_t,b_t }\}^k_{t=1}$. Specifically, $F_K$ can be generated by using the keyed-hash message authentication code $HMAC$, i.e., $F_K=HMAC(K,\cdot)$ and $K$ is a randomly generated secret key of length $\lambda$. $H_1$ can be generated using the random oracle $H$.

\subsection{IndexGen}\label{5-2}

\begin{figure}
\centering
\includegraphics[width=3in]{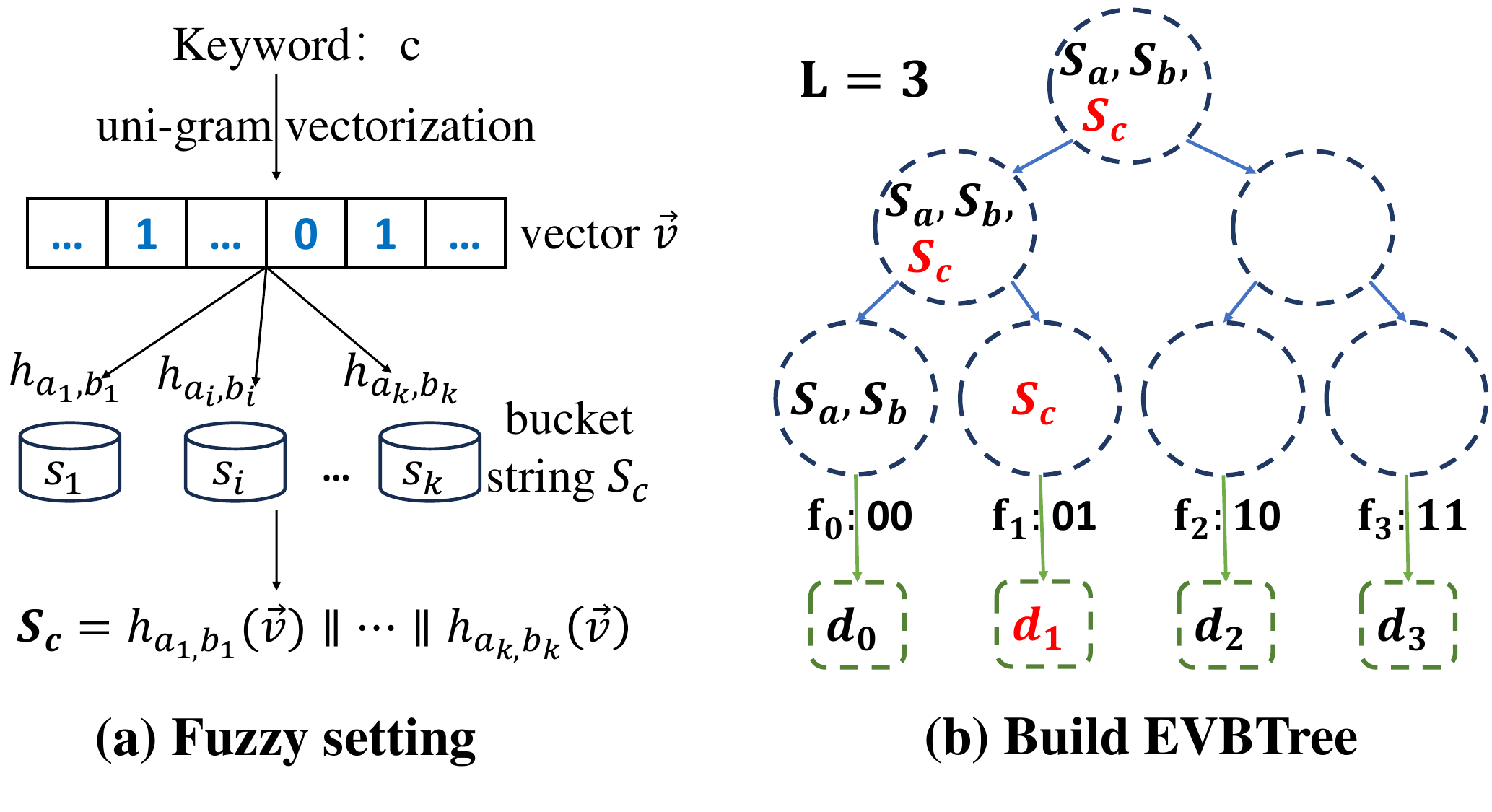}
\caption{An instance of the EVBTree with a height of $L = 3$ and four leaves. } \label{evbtree}
\end{figure}

To achieve the fault tolerance requirement for user queries in \textbf{G1}, given an outsourced file set $F = \{f_1,f_2,\cdots,f_n\}$, the DO extracts keywords from each record. Applying a stemming algorithm (e.g., the Porter Stemming Algorithm \cite{porter2006algorithm}), the DO derives keyword dictionaries $W(f_i)$ by reducing related words (e.g., ``encrypted", ``encrypting", and ``encrypts" to ``encrypt"). Then, as shown in Fig.~\ref{evbtree}, the DO generates an encrypted logical index tree, EVBTree. Algorithm \ref{alg:index_gen} provides a detailed explanation of index generation.

\begin{algorithm}[!ht]
\caption{Secure and Fuzzy Index Construction (EVBTree)}
\label{alg:index_gen}
\begin{algorithmic}[1]
\REQUIRE  Document collection $D = \{f_1, f_2, \dots, f_n\}$, master secret key $K$, LSH family $LSH_k = \{h_1, h_2, \dots, h_k\}$, cryptographic hash functions $H_1$, $F_K$
\ENSURE Secure search index $I$ (implemented as a hash table)

\STATE Initialize an empty hash table $I$ \COMMENT{The logical EVBTree}

\FOR{each document $f_i \in D$}
    \STATE $W_i \gets \text{ExtractKeywords}(f_i)$ \COMMENT{Apply stemming, stop-word removal, etc.}
    \STATE $leafPath \gets \text{GetBinaryPath}(i)$ \COMMENT{e.g., for $i=3$, $path$=`11' for a height-3 tree}
    \STATE $d_i \gets F_K(leafPath \| \text{Enc}_K(f_i))$ \COMMENT{Compute document digest}
    \STATE Store $\langle \text{`digest'}\|leafPath, d_i \rangle$ in $I$ \COMMENT{Store digest for future verification}
    
    \FOR{each keyword $w \in W_i$}
        \STATE $\vec{v} \gets \text{UnigramVector}(w)$ \COMMENT{Convert $w$ to a 160-dim binary vector}
        \STATE $S_w \gets \emptyset$
        \FOR{each LSH function $h_t \in LSH_k$}
            \STATE $bucket \gets h_t(\vec{v})$ \COMMENT{Compute the LSH bucket ID for $w$ under $h_t$}
            \STATE $S_w \gets S_w \cup \{bucket\}$ \COMMENT{$S_w$ is the set of all LSH buckets for $w$}
        \ENDFOR
        
        \STATE $nodeList \gets \text{GetNodes}(leafPath)$ \COMMENT{Get all nodes on the path to $f_i$'s leaf}
        \FOR{each node $v \in nodeList$}
            \STATE $nodePath \gets \text{GetBinaryPath}(v)$ \COMMENT{e.g., root=``", left-child=``0", right-child=``01"}
            \FOR{each $bucket \in S_w$}
                \STATE $key \gets H_1(nodePath \| F_K(bucket))$ \COMMENT{Generate the unique key for the hash table $I$}
                \STATE $value \gets 1$ \COMMENT{Value can be a counter or a flag indicating presence}
                \STATE Store $\langle key, value \rangle$ in $I$ 
            \ENDFOR
        \ENDFOR
    \ENDFOR
\ENDFOR
\RETURN $I$
\end{algorithmic}
\end{algorithm}

\begin{itemize}
    
\item \textit{Fuzzy Setting}. We employ the vectorization algorithm based on the uni-gram and locality-sensitive hashing (LSH) technique to process keywords, enhancing the fault tolerance of user queries. Specifically, for each keyword $w_j\in W(f_i)$, DO first converts $w_j$ into a 160-dimensional uni-gram vector $\vec{v_j}$, which serves as the input for LSH. Subsequently, LSH family $LSH_k=\{h_{a_t,b_t}\}_{t=1}^k$ are utilized to compute the the key-value pairs $WD(w_j, i) = \{(H_1 (path(v) \| F_K (LSH_k(\vec{v_j}) )), 1)\}_{v \in nodes(leaf_i)}\label{T1}$, where \(path(v)\) is the string path encoding of node \(v\), used to identify the node position in the virtual binary tree.

\item \textit{Index EVBTree Built}. Given a file $f_i$ with a keyword set $W(f_i )=\{w_1,w_2,\cdots\}$, the application of a fuzzy setting to this keyword set results in a bucket string value set $\{S_{w_1},S_{w_2},\cdots\}$. Additionally, according to the path encoding rules of the VBTree structure, the path value $path(f_i)$ of the file $f_i$ is a string composed of 0 and 1, numerically representing the binary form of $i$. For each bucket string $S_{w_j}$, we calculate insert value $t_{S_{w_j}}=\{H_1 (path(v)\| F_K (S_{w_j} ))\}_{v\in nodes(leaf_i)}$. Here, node $v$ is a traversal node from the root to the leaf node $leaf_i$ of the document $f_i$ according to the design of vbtree structure. The set $nodes(leaf_i)$ comprises the traversal nodes, and $path(v) $ denotes the string path code of node $v$. Subsequently, the digest of a file $f_i$ is $d_i=F_K (path(leaf_i)\|c_i )$ is inserted into the index EVBTree $I$, where \(path(leaf_i)\) denotes the path encoding information of leaf node $leaf_i$ corresponding to document identifier $i$, and $c_i$ is the encrypted document $f_i$.. Finally, all the insert values and digests are stored in the EVBTree I, which logically adopts a binary tree structure but is actually implemented as a hashtable. The details of building an index EVBTree are shown in Algorithm \ref{alg:index_gen}.
\end{itemize}

The \textbf{time complexity of the IndexGen Algorithm \ref{alg:index_gen}} is analyzed as follows:

Let $n$ be the number of documents, $|W_i|$ be the average number of keywords per document, and $k$ be the number of LSH functions. For each document $f_i$, keyword extraction is $O(|W_i|)$, then for each keyword, LSH processing requires $O(k)$ operations, and for each LSH bucket, VBTree path insertion requires $O(L)$ operations, where $L$ is the tree height ($L = \lceil \log_2 n \rceil$).
Thus, the total complexity is:
\[
T(n) = O\left(n \cdot |W_i| \cdot k \cdot L\right) = O\left(n \cdot |W_i| \cdot k \cdot \log n\right)
\]
Since $|W_i|$ and $k$ are constants independent of $n$, the complexity simplifies to $T(n) = O(n \log n)$, confirming our claim of logarithmic scaling per document insertion during initial index construction.

\subsection{Update}\label{5-3}
To achieve the goals of supporting efficient file addition (\textbf{G1}) while ensuring forward privacy (\textbf{G4}), we design a novel dynamic update mechanism based on dual repositories (Local Repository and Blockchain Repository) and the EVBTree structure. 
Algorithm \ref{alg:document_addition} details the main document addition process. Algorithm \ref{alg:update_keyword_state} focuses on the keyword status update section, which is the core of achieving forward security. Algorithm \ref{alg:document_deletion} describes the document deletion mechanism.

\begin{algorithm}[!ht]
\caption{Document Addition and Index Update}
\label{alg:document_addition}
\begin{algorithmic}[1]
\REQUIRE New document set $F = \{f_1, f_2, \dots, f_m\}$ to add, master secret key $K$, local repository (client-side) $LR$, blockchain repository (on-chain) $BR$ and existing EVBTree index (on-chain) $I$
\ENSURE Updated index $I$, updated repositories $LR$ and $BR$

\FOR{each new document $f_i \in F$}
    \STATE \Com{// Phase 1: Document Processing}
    \STATE $W_i \gets \text{ExtractKeywords}(f_i)$ \COMMENT{Apply stemming, stop-word removal}
    \STATE $leafPath_i \gets \text{GetBinaryPath}(i)$ \COMMENT{Determine VBT leaf path for doc ID $i$}
    \STATE $c_i \gets \text{Enc}_K(f_i)$ \COMMENT{Encrypt document content}
    \STATE $\text{StoreCiphertext}(leafPath_i, c_i)$ \COMMENT{Store on cloud storage}
    
    \STATE \Com{// Phase 2: Index Construction for New Document}
    \STATE $IndexEntries_i \gets \emptyset$
    \FOR{each keyword $w \in W_i$}
        \STATE \Com{// Process keyword for fuzzy search}
        \STATE $\vec{v}_w \gets \text{UnigramVector}(w)$ \COMMENT{Convert to 160-dim binary vector}
        \STATE $S_w \gets \{ h_t(\vec{v}_w) \mid h_t \in LSH_k \}$ \COMMENT{Generate LSH buckets}
        
        \STATE \Com{// Generate index entries for EVBTree}
        \STATE $nodeList \gets \text{GetNodes}(leafPath_i)$
        \FOR{each node $v \in nodeList$}
            \STATE $nodePath \gets \text{GetBinaryPath}(v)$
            \FOR{each bucket $b \in S_w$}
                \STATE $indexKey \gets H_1(nodePath \| F_K(b))$
                \STATE $IndexEntries_i \gets IndexEntries_i \cup \{ (indexKey, 1) \}$
            \ENDFOR
        \ENDFOR
        
        \STATE \Com{// Update keyword state for forward privacy}
        \STATE UpdateKeywordState($w$, $i$, $LR$, $BR$) \COMMENT{Call Alogrithm \ref{alg:update_keyword_state}}
    \ENDFOR
    
    \STATE \Com{// Phase 3: Blockchain Update}
    \STATE $\text{SubmitToBlockchain}(IndexEntries_i)$ \COMMENT{Add to on-chain index $I$}
\ENDFOR
\end{algorithmic}
\end{algorithm}

\begin{algorithm}[!ht]
\caption{UpdateKeywordState($w$, $m$, $LR$, $BR$)}
\label{alg:update_keyword_state}
\begin{algorithmic}[1]
\REQUIRE Keyword $w$ being updated, document ID $m$ containing the keyword, local repository $LR$ and blockchain repository $BR$
\ENSURE Updated $LR$ and $BR$

\IF{$LR[w].b == \textsf{false}$} 
    \STATE \Com{// First update for this keyword session}
    \STATE $T_{\text{add}} \gets WD(w, m) \setminus WD(w, LR[w].n_{\text{add}})$
    \STATE $T_{\text{add}} \gets \text{PadToLength}(T_{\text{add}}, L)$ \COMMENT{Privacy-preserving padding}
    \STATE $I \gets I \cup T_{\text{add}}$ \COMMENT{Add delta to main index}
\ELSE
    \STATE \Com{// Subsequent update in ongoing session}
    \STATE $v \gets LR[w].v$
    \STATE $key_{v} \gets H_2(F_K(w \| v))$
    \STATE $value_{v} \gets H_3(F_K(w \| v)) \oplus F_K(w \| (v-1))$
    \STATE $T_{\text{chain}} \gets (key_{v}, value_{v})$
    \STATE $BR \gets BR \cup T_{\text{chain}}$ \COMMENT{Store chaining element}
\ENDIF

\STATE \Com{// Update local repository state}
\STATE $LR[w].v \gets LR[w].v + 1$
\STATE $LR[w].b \gets \textsf{true}$
\STATE $LR[w].n_{\text{add}} \gets m$
\end{algorithmic}
\end{algorithm}

\begin{algorithm}[!ht]
\caption{Document Deletion}
\label{alg:document_deletion}
\begin{algorithmic}[1]
\REQUIRE Keyword $w$ to remove, document ID $m$ from which to remove the keyword, deletion logs $DL$ (on-chain) and local repository $LR$
\ENSURE Updated $DL$ and $LR$

\STATE \Com{// Calculate the index entry to remove}
\STATE $nodeList \gets$ \text{GetNodes}(\text{GetBinaryPath}$(m)$)
\STATE $\vec{v}_w \gets \text{UnigramVector}(w)$
\STATE $S_w \gets \{ h_t(\vec{v}_w) \mid h_t \in LSH_k \}$
\STATE $entry_{\text{del}} \gets \emptyset$

\FOR{each node $v \in nodeList$}
    \STATE $nodePath \gets \text{GetBinaryPath}(v)$
    \FOR{each bucket $b \in S_w$}
        \STATE $key \gets H_1(nodePath \| F_K(b))$
        \STATE $entry_{\text{del}} \gets entry_{\text{del}} \cup \{ key \}$
    \ENDFOR
\ENDFOR

\STATE \Com{// Mark for deletion instead of immediate removal}
\STATE $DL \gets DL \cup entry_{\text{del}}$
\STATE $LR[w].n_{\text{del}} \gets m$ \COMMENT{Record deletion in local repository}
\end{algorithmic}
\end{algorithm}

Our update mechanism achieves the following properties:
\begin{itemize}
\item \textbf{Security}: The dual-repository design (Local Repository and Blockchain Repository in Algorithm \ref{alg:update_keyword_state}) enables efficient state management while maintaining strong security guarantees. The Local Repository keeps small-sized state information, while the Blockchain Repository provides tamper-proof storage for update operations.

\item \textbf{Forward Privacy (\textbf{G4})}: Through the Algorithm \ref{alg:update_keyword_state}, we ensure that new additions cannot be linked to previous search operations. The cryptographic chaining mechanism in $BR$ prevents correlation between different versions of the same keyword.

\textbf{Efficiency (\textbf{G1})}: By leveraging the EVBTree structure, updates require only $O(L \cdot |W_i|)$ operations (as shown in Algorithm \ref{alg:document_addition}), where $L$ is the tree height, which is significantly better than the $O(n)$ complexity of linear schemes.

\item \textbf{Verifiability}: All update operations are recorded on-chain through the Blockchain Repository, enabling transparent audit trails and verification of update correctness.

\item \textbf{Backward Privacy}: The deletion mechanism using a Deletion Journal (in  Algorithm \ref{alg:document_deletion}) ensures that removed documents do not appear in search results while preventing the server from learning which specific documents were deleted.
\end{itemize}

The \textbf{time complexity of the Update Algorithm \ref{alg:document_addition}} is analyzed as follows:

For each new document $f_i$, $O(|W_i|)$ for keyword extraction and vectorization, and $O(|W_i| \cdot k)$ for generating LSH buckets, then each keyword requires insertion along the VBTree path of length $L = O(\log n)$.
The overall complexity for adding one document is:
\[
T_{\text{add}}(n) = O(|W_i| \cdot k \cdot \log n) = O(\log n)
\]

This logarithmic complexity demonstrates that our semantic-decoupled indexing successfully avoids the $O(n^2)$ update overhead that plagued previous integrated approaches like Zhong et al. \cite{Zhong2020Efficient}.

\subsection{TrapGen}\label{5-4}
To search the encrypted index, DU formulates a search trapdoor with encrypted conditions from user provided keywords and sends it to BC.
For a query $Q = \{w_1',w_2',\cdots,w_q' \}$ of $q$ keywords, each $w_i'\in Q$ is first vectorized to $\vec{v_i'}$ via a uni-gram algorithm. Then, set $LR(w_i).b$  in the local repository to $true$ to mark that the keyword has been queried, for each $\vec{v_i'}$, an LSH function calculates a bucket string $S_{w_i'}$ as per Eq.~\eqref{eq1}. 
\begin{equation}
          S_{w_i'}=LSH_k (\vec{v_i'})=h_{a_1,b_1 } (\vec{v_i'} )\|\cdots\|h_{a_k,b_k } (\vec{v_i'} )    \label{eq1}
\end{equation}
Next, each bucket string $S_{w_i'}$ undergoes encryption using a pseudo random function $F_K$ to generate a token $tk_i$,combined with the current version number $LR(w_i).v$ of the keyword $w_i$, as dictated by the Eq.~\eqref{eq2}:
\begin{equation}
          tk_i = F_K(LSH_k(\vec{v_i}) || LR(w_i).v)  \label{eq2}
\end{equation}
Upon completion of the processing for all query keywords, the search trapdoor $TK=\{tk_1,\cdots,tk_q\}$ is constructed, where $q$ denotes the total number of query keywords in $Q$.

\subsection{Search}\label{5-5}

Algorithm~\ref{alg:search} presents the core search procedure of VeriFuzzy, which achieves \textbf{G3} by enabling efficient fuzzy multi-keyword search with $O(\log n)$ complexity through the EVBTree structure.
The VeriFuzzy search protocol executes in three sequential phases as follows:
\begin{itemize}
\item \textit{Phase 1- Historical Trapdoor Preparation}: For each keyword $w_i$ in query $Q = \{w_1, \dots, w_q\}$, the system generates the current trapdoor $tk_i = F_K(\text{LSH}_k(\vec{v_i}) \parallel \text{LR}(w_i).v)$. The blockchain repository $BR$ is queried to retrieve all historical trapdoors $h(w_i) = \{x_{i1}, \dots, x_{ip_i}\}$ where $p_i$ represents the version count. The complete historical trapdoor set $H = \{h(w_1), \dots, h(w_q)\}$ is constructed for the search operation.
\item \textit{Phase 2- Recursive Index Traversal}: The search begins at the root node (empty path) and recursively traverses the EVBTree structure. For each node, the algorithm verifies if \textit{all} query keywords have at least one historical version present by checking $H_1(path \| x_{ij})$ for each $x_{ij} \in h(w_i)$. Branches are pruned early when any keyword lacks matching entries, significantly improving search efficiency. Matching nodes are recorded in auxiliary proof set $AP$ with their path and match status (0/1). Leaf nodes reached through successful paths are added to the result set $R$ with their corresponding digests $d_i$ stored for verification.
\item \textit{Phase 3- Result Filtering and Verification}: The preliminary result set $R$ is filtered against the deletion logs $DL$ to remove any documents that have been marked for deletion. The final result set $R_{\text{final}}$ is returned along with auxiliary proofs $AP$ and digests $D$ for blockchain-based verification. The entire process maintains $O(\log n)$ complexity due to the tree structure and early pruning mechanism.
\end{itemize}

The \textbf{time complexity of the Search Algorithm \ref{alg:search}} is analyzed considering the worst-case scenario:

Let $q$ be the number of query keywords, $p$ be the maximum historical versions per keyword, and $L$ be the VBTree height. $O(q \cdot p)$ for retrieving all versioned trapdoors, and in worst case, we traverse $O(2^L) = O(n)$ nodes, but our early pruning strategy ensures that each node requires $O(q \cdot p)$ comparisons, and with effective pruning, only $O(L) = O(\log n)$ nodes are visited in practice. In deletion filtering phase needs $O(|R|)$ where $|R|$ is the result set size.

The dominant term is the tree traversal:
\[
T_{\text{search}}(n) = O(q \cdot p \cdot \log n) = O(\log n)
\]

In summary, we have made three optimizations for achieving efficient retrieval.
First, we ensure that documents remain searchable after multiple updates while maintaining forward security by checking all historical versions of each keyword. Second, we significantly reduce the number of index accesses by immediately terminating the recursive search when any keyword mismatch occurs. Finally, instead of continuously maintaining deleted indexes, we perform deletion filtering after searching to improve update efficiency while ensuring accuracy. Our time complexity analysis confirms the $O(\log n)$ search complexity claimed in our introduction, representing a significant improvement over linear-scan approaches.

\begin{algorithm}[!ht]
\caption{EVBTree Search Protocol}
\label{alg:search}
\begin{algorithmic}[1]
\REQUIRE Encrypted EVBTree index $I$, full historical trapdoor set $H = \{h(w_1), h(w_2), \dots, h(w_q)\}$, current node path $path$ (initially empty string for root) and deletion logs $DL$
\ENSURE Result set $R$, Auxiliary proofs $AP$, Digest set $D$

\STATE \Com{// Phase 1: Check all keyword versions at current node}
\FOR{each keyword $w_i \in Q$}
    \STATE $keyword\_matched \gets \textsf{false}$
    \FOR{each historical version $x_{ij} \in h(w_i)$}
        \STATE $key \gets H_1(path \| x_{ij})$
        \IF{$I.\text{contains}(key)$}
            \STATE $keyword\_matched \gets \textsf{true}$
            \STATE \textbf{break} \COMMENT{One matching version is sufficient}
        \ENDIF
    \ENDFOR
    
    \IF{\NOT $keyword\_matched$}
        \STATE \Com{// Current node doesn't contain all keywords - prune this branch}
        \STATE $AP \gets AP \cup \{(path, 0)\}$ \COMMENT{Record mismatch for verification}
        \RETURN $\emptyset$ \COMMENT{Prune this search branch}
    \ENDIF
\ENDFOR

\STATE \Com{// Phase 2: All keywords matched at this node - continue processing}
\STATE $AP \gets AP \cup \{(path, 1)\}$ \COMMENT{Record match for verification}

\IF{$\text{isLeaf}(path)$}
    \STATE \Com{// Reached a leaf node - add to potential results}
    \STATE $id \gets \text{BinaryToDecimal}(path)$
    \STATE $R \gets R \cup \{id\}$
    \STATE $D \gets D \cup \{d_{id}\}$ \COMMENT{Store digest for later verification}
\ELSE
    \STATE \Com{// Continue searching in both subtrees}
    \STATE $R_{\text{left}} \gets \text{Search}(I, H, path\|'0', DL)$
    \STATE $R_{\text{right}} \gets \text{Search}(I, H, path\|'1', DL)$
    \STATE $R \gets R \cup R_{\text{left}} \cup R_{\text{right}}$
\ENDIF

\STATE \Com{// Phase 3: Post-processing - apply deletion filtering}
\STATE $R_{\text{final}} \gets R$
\FOR{each document $id \in R$}
    \IF{$\text{IsDeleted}(id, DL, H)$}
        \STATE $R_{\text{final}} \gets R_{\text{final}} \setminus \{id\}$
    \ENDIF
\ENDFOR

\RETURN $R_{\text{final}}$
\end{algorithmic}
\end{algorithm}

\subsection{Verify}\label{5-6}
\begin{figure}
\centering
\includegraphics[width=3in]{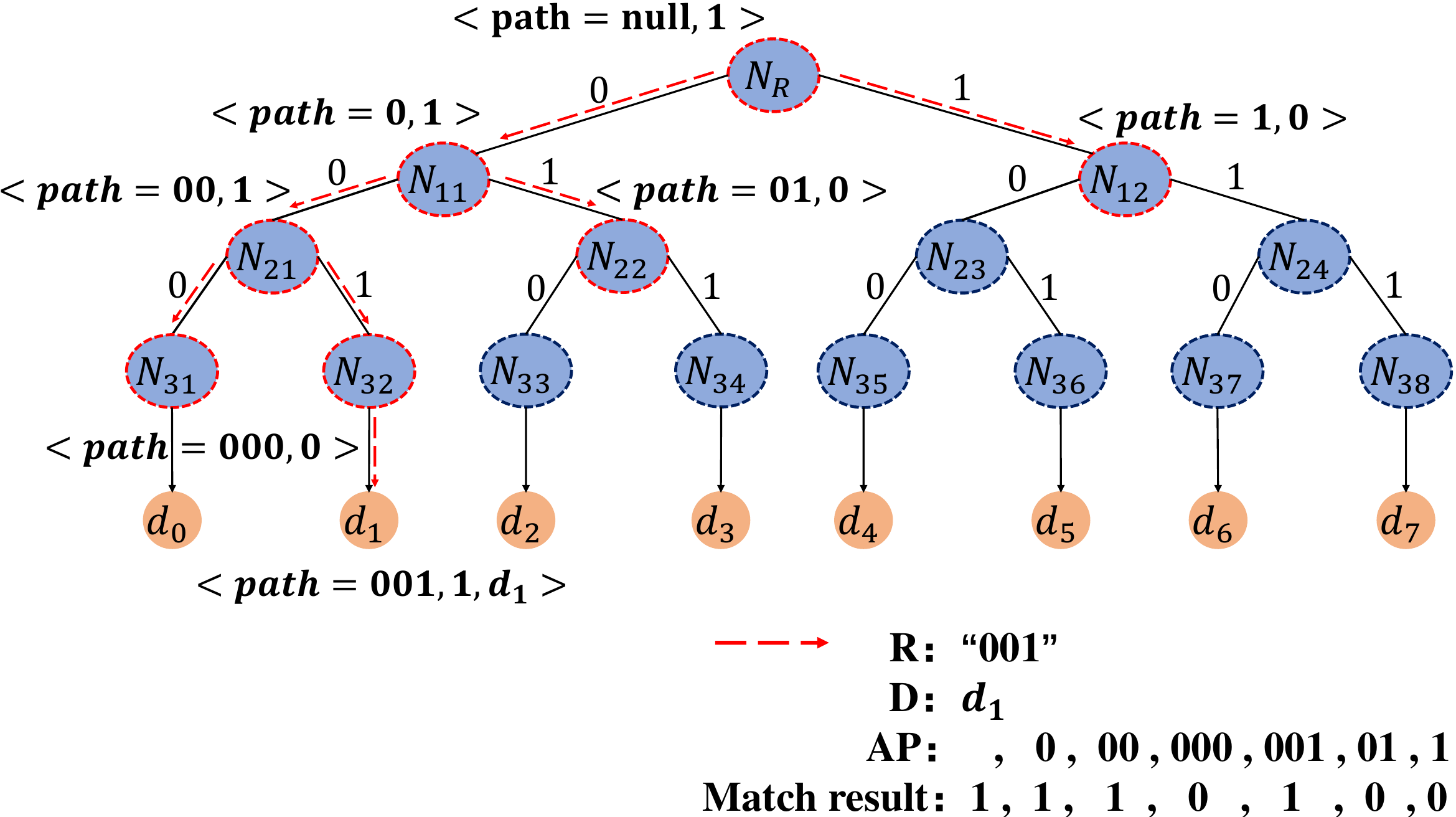}
\caption{Example of the auxiliary proof information for the search results $c_1$.} \label{fig4}
\end{figure}

Upon receiving the result set $R$, the auxiliary proof set $AP$, the ciphertext digests set $D$, and the ciphertext set $E$, the smart contract verifies the correctness and completeness of the search results. The design not only accomplishes \textbf{G2} but also achieves the super-linear verification efficiency in \textbf{G3}.

Specially, first, verification of result correctness. Given a result set R, comprising r search results denoted as $R=\{path(leaf_0 ) ,\cdots,path(leaf_r )\}$, and the corresponding encrypted files $E=\{c_0 ,\cdots,c_r\}$ retrieved from the cloud server, the contract recomputes the digests set 
\begin{equation}
\begin{aligned}
    D'& =\{d_0',d_1',\cdots,d_r'\} \\
    & = \{F_K (path(leaf_0 )\|c_0),
\cdots,F_K (path(leaf_r )\|c_r)\} 
\end{aligned}
\end{equation}
for each leaf node suffix and compares them against the search-derived set $D=\{d_0,d_1,\cdots,d_r\}$, verifying whether the equation $D'=D$ holds. The comparison outcome serves as evidence for the tampering status of the stored records. Secondly, verification of Result Integrity. Upon correct verification, the verify contract uses $AP$ to build a search binary tree via node path encoding and matching. It checks if non-leaf terminal node search results are 0; if yes, it negates child node matches, skipping downward search and passing verification. Otherwise, non-zero results imply incomplete ciphertext retrieval, missing potential matches. If all verifications succeed, the smart contract returns 1 to the user, and the cloud server provides the ciphertext data.

The \textbf{time complexity of our blockchain-based verification} through path reconstruction:

The smart contract reconstructs only the search path of length $L = O(\log n)$, and by Merkle tree in blockchian, each node on the path requires $O(1)$ hash operations, then $O(|R|)$ for result set verification, where $|R|$ is typically small.
So the overall verification complexity is:
\[
T_{\text{verify}}(n) = O(\log n + |R|) = O(\log n)
\]


For better understanding, let's explain it through Fig.~\ref{fig4}. Given query $Q = \{w_1, w_2\}$ for example, assume only $f_1$ among $\{f_0,f_1,f_2,f_3,f_4,f_5,f_6,f_7\}$ has both keywords. Then, $path(N_{32})=$``001'' is in search result $R$, and ciphertext digest $d_2$ is in $D$. The auxiliary proof set $AP = \{$``'',``0'',``1'',``00'',``01'',``000'',``001''$\}$ contains traversed node path encodings, with matching results $\{1,1,0,1,0,0,1\}$ ($1$: match, $0$: no match). In verification, recalculate $d_1'$ per Eq.~\eqref{eq4} and compare with $d_1$. Since $d_1' = d_i$, the ciphertext result is correct. Then, construct a search binary tree based on search paths and node matching results (Fig.~\ref{fig4}). As non-leaf terminal node search results are 0, the search-path integrity is validated, with no unqueried nodes missing. 
\begin{equation}
          d_1'=F_K (path(N_{32} )\|c_1 )=F_K ('001'\|c_1 )   \label{eq4}
\end{equation}

\section{Security Analysis}\label{SecurityAnalysis}
Based on the security definition, we prove that our scheme is IND-CKA2($L_1,L_2,L_3$)-secure in Theorem \ref{theorem1}, achieving objective \textbf{G4}.

\begin{theorem}\label{theorem1}
If $F_K$ is a pseudo-random function and $H_1$ is a random oracle, then our scheme is IND-CKA2($L_1,L_2,L_3$)-secure against an adaptive adversary.
\end{theorem}

\begin{proof}
We describe a simulator S such that for any PPT adversary $\mathcal{A}$, the outputs of $Real_{\mathcal{A}}(\lambda)$ and $Sim_{\mathcal{A},S}(\lambda)$ are computationally indistinguishable. 
The simulator $S$ adaptively generates the index and search trapdoors as follows:

$S(\lambda,L_1(F))$: $S$ simulates the ciphertext set $C'=\{c_1',\cdots,c_n'\}$, whose security is guaranteed by the CPA-secure encryption algorithm, along with the file identifiers and the size of each ciphertext, which are included in the leakage function $L_1(F)$. $S$ then creates a simulated hash table, designated as the generated simulated index $I'$, to accommodate the tree, which is padded with random values to the same size as the original hash table. Finally, $S$ sends $I'$ to $A$.
$S(L_2(F,Q_1))$: After receiving a search query $Q_1=\{F_K(w_{11}),F_K(w_{12}),\cdots,F_K(w_{1u})\}$ from $\mathcal{A}$, $S$ generates $u$ random values $\{z_{11},z_{12},\cdots,z_{1u}\}$ with the same size as the original trapdoor values as the simulated trapdoor such that
\begin{equation}
\begin{aligned}
    & d_1'=F_K (path(N_{32} )\|c_1 ) \\
    & =F_K F(z_{11})\cap \cdots \cap F(z_{1u} )=R_{Q_1}   \label{eq5}
\end{aligned}  
\end{equation}
where $R_{Q_1}$ is the search results revealed by the leakage function $L_2(F,Q_1)$, and $F(z_{1j})(1\le j\le u)$ represents the set of documents retrieved as search results via $z_{1j}$, wherein for each node $v$ traversed along the path from the root node to the respective leaf node corresponding to the result, the computed hash $H_1(path(v)\|z_{1j})$ is included. $S$ then sends the simulated trapdoor $TK_1'=\{z_{11},z_{12},\cdots,z_{1u}\}$ to $A$. 

$S(L_2(F,Q_1,\cdots,Q_i))$: Assuming that $\mathcal{A}$ searches the index $s$ times, the simulator $S$ simulates the search trapdoors $s$ times. Upon receiving a search query $Q_i=\{F_K (w_{i1}),F_K (w_{i2}),\cdots,F_K (w_{iu})\}(1\le i\le s)$, $ \{F_K (w_{i1}),\\F_K (w_{i2}),\cdots,F_K (w_{ix})\}(1\le x\le u)$ have appeared before according to the leakage function$ L_2 (F,Q_1,\cdots,Q_i )$, thereby $S$ uses these values $\{z_{i1},z_{i2},\cdots,z_{ix}\}$ for simulation. For the remaining $u-x$ terms, $S$ generates $u-x$ never-used random values $\{z_{ix+1},z_{ix+2},\cdots,z_{iu}\}$ with the same size as the original trapdoor values as the simulated trapdoor such that
\begin{equation}
          F(z_{i1} )\cap \cdots \cap F(z_{ix})\cap F(z_{ix+1}))\cap\cdots\cap F(z_{iu})=R_{Q_i}   \label{eq6}
\end{equation}
where $R_{Q_i}$ is the search results revealed by the leakage function $L_2 (F,Q_1,\cdots,Q_i )$, $F(z_{iy})(1\le y\le u)$ represents the set of documents retrieved as search results via $z_{iy}$, wherein for each node $v$ traversed along the path from the root node to the respective leaf node corresponding to the result, the computed hash $H_1 (path(v)\|z_{iy} )$ is included. $S$ then sends the simulated trapdoor $TK_i'=\{z_{i1}  ,z_{i2}  ,\cdots,z_{ix},z_{ix+1},z_{ix+2} ,\cdots,z_{iu}\}$ to $A$.

$S(L_3(W,id))$: If $\mathcal{A}$ updates the index $s$ times, the simulator $S$ simulates updating the index $s$ times. After receiving an additional keyword set $W_j=\{w_1,w_2,\cdots,w_m\}(1\le j\le k)$, $S$ generates m random values $\{r_1,r_2,\cdots,r_m\}$ of the same size as $H_1 (path(v)\|F_K (w_n))(1\le n\le m)$ to serve as simulated insertion values.  These values are then appended to form an updated index $I_{up}'$ which is sent to $A$.

After that, $\mathcal{A}$ attempts to distinguish the results $\theta=\{I,I_{up},\{TK_1,\cdots,TK_s\}\}$ returned by the challenger from the outputs $\theta'=\{I',I_{up}',\{TK_1',\cdots,TK_s'\}\}$ returned by the simulator $S$. Then, we have
\begin{equation}
\begin{aligned}
    & |\Pr[D(\theta)=1]-\Pr[D(\theta')=1]| \\
    & \leq Adv_1+Adv_2+Adv_3 \label{eq7}
\end{aligned}
\end{equation}
where $Adv_1$ is the advantage of $A$ to distinguish $I$ from $I'$; $Adv_2$ is the advantage of $A$ to distinguish $I_{up}$ from $I_{up}'$; $Adv_3$ is the advantage of $A$ to distinguish $\{TK_1,\cdots,TK_s \}$ from $\{TK_1',\cdots,TK_s'\}$.

From the simulation process, we know that $Adv_1$ represents the advantage in distinguishing the pseudo-random function $F_K$ from a real random function. The scheme \cite{r21} has demonstrated that a pseudo-random function is computationally indistinguishable from a random function. Moreover, each simulated ciphertext $C'$ is indistinguishable from a real ciphertext $C$ as the ciphertext is encrypted using AES, which is CPA-secure if $A$ does not know the encryption key. Thus, $Adv_1\le negl_1 (\lambda)$. $Adv_2$ and $Adv_3$ are also expressed as the distinction of the outputs of the pseudo-random functions $F_K$ and real random functions. For the same reason, $Adv_2\le negl_2 (\lambda)$, $Adv_3\le negl_3 (\lambda)$.
\begin{equation}
\begin{aligned}
    & |Pr[D(\theta)=1)]-Pr[D(\theta')=1]| \\
    & \le negl_1 (\lambda)+negl_2(\lambda)+negl_3(\lambda) \\
    & = negl(\lambda)\label{eq8}
\end{aligned}
\end{equation}
Therefore, we can conclude that our scheme is secure under the IND-CKA2 model.
\end{proof}

\section{Performance Evaluation}\label{sec:performance}

In this section, we conduct a systematic evaluation of the core functionalities of the proposed scheme through comprehensive experiments. The experiments are primarily organized into three aspects: For Robustness, we test fuzzy search effectiveness in handling spelling errors and evaluate the time complexity of dynamic document addition. For efficiency, we measure the time overhead of the search algorithm and compare its performance with existing solutions. For verifiability, we analyze the time complexity of the verification functionality and compare it with other schemes. Based on the experimental results, we address the following key research questions:

\begin{itemize}
    \item \textbf{RQ1:} How effective is the proposed scheme in handling spelling errors under the fuzzy setting, and what is the time complexity of the dynamic document addition functionality?
    \item \textbf{RQ2:} How does the time overhead of the search functionality in our scheme compare to existing solutions, and how much is the search efficiency improved?
    \item \textbf{RQ3:} What is the computational overhead of our verification algorithm compared to other scheme?
\end{itemize}

\subsection{ Experiment Setup and Baselines}\label{AA}
This section systematically evaluates the core functions of VeriFuzzy in terms of \textbf{robustness}, \textbf{efficiency}, and \textbf{verifiability}. 
Our implementation of the cryptographic algorithms and the searchable encryption scheme is publicly available at: \url{https://github.com/JackAugust/VeriFuzzy.git}. Note that the blockchain smart contract components are excluded due to commercial deployment constraints.

\textbf{DataSets} We select two datasets to evaluate our scheme. One dataset is the RFC (request for comments) dataset with terminology-dense characteristics, from which 56,658 keywords were extracted\footnote{By using RAKE algorithm. \url{https://github.com/aneesha/RAKE}} from 3,500 files, with the number of keywords per file about 92. The other dataset is the NSF Research Award abstracts spanning 1990 to 2003\footnote{The dataset can be found at http://kdd.ics.uci.edu/databases/nsfabs/nsfawards.html} with term sparsity characteristics, from which 3,167 keywords were extracted from 500 documents, with an average of approximately 45 keywords per file.

\textbf{Implementation details} We conduct our experiments mainly on a PC running Windows 10 with a 1.60GHz Intel(R) Core(TM) i5-8250U CPU and 8GB memory, using the Java Pairing-Based Cryptography library. We set the number of HMAC-SHA256 functions to $k=8$ as the pseudorandom function and use SHA-512 as the random oracle. We employ a Bloom filter of length $N = 1000$, resulting in an average false positive rate of less than $(1-e^{-|W|\times k/N})^k=(1-e^{-92\times 8/1000})^8 \approx 0.5\%$. Using the LSH parameters $(\sqrt{3},\ 2,\ 0.56,\ 0.28)$. The height of the VBTree $l=32$.
Our blockchain environment is under a cloud server running Ubuntu 20.04 64-bit operating system with Hyperledger Fabric V2.2, developing the contract described in this paper based on Go 1.18.8.

\textbf{Baselines.} As shown in Table \ref{tab:comprehensive-comp}, to comprehensively evaluate the performance of the fuzzy search, dynamic update, and result verification algorithms in our VeriFuzzy, we compare its performance with several existing solutions. For fuzzy search, we compared VeriFuzzy with DVFKF\cite{cuidynamic2025}, VFSA\cite{tong2023verifiable}, PCQS\cite{li2017adaptively}, and PCKS-FS\cite{Liutowards2025} (as shown in Fig.~\ref{Evaluation} (a, d, e)). For dynamic update, we compared VeriFuzzy with DVFKF\cite {cuidynamic2025}, EDMF\cite{Zhong2020Efficient} for dynamic updates (as shown in Fig.~\ref{Evaluation} (b,c)), and with DVFKF\cite{cuidynamic2025}, VFSA\cite{tong2023verifiable}, MVSSE\cite{liu2020multi}, and MVDSE\cite{xuforward2025} for verification (as shown in Fig.~\ref{Evaluation} (f)).

\subsection{Evaluation of Fuzzy Setting and Addition\textbf{ (RQ1)}}\label{RQ1}

\begin{figure*}[htp]
\centering
\includegraphics[width=1\textwidth]{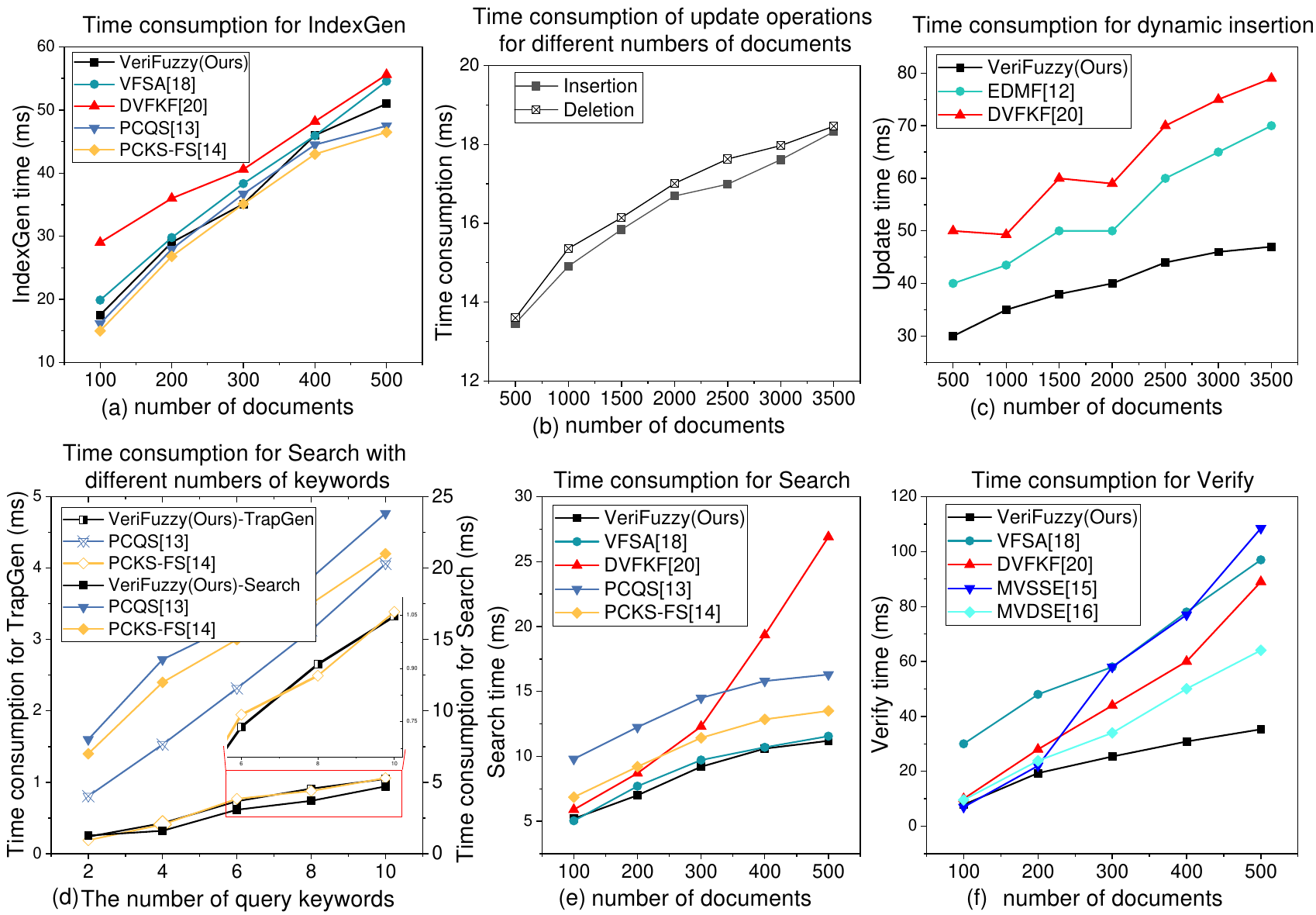}
\caption{Performance comparison between our VeriFuzzy and existing solutions. Time consumption for \textbf{IndexGen}(Fig.a), \textbf{Update}(Fig.b,c), \textbf{TrapGen}(Fig.d), \textbf{Search}(Fig.e), and \textbf{Verify}(Fig.f).}
\label{Evaluation}
\end{figure*}

\textbf{1) Fuzzy Setting:} We evaluate the effectiveness of the fuzzy setting in handling spelling errors through search accuracy. We categorize spelling errors in keywords into four common types: misspelling a letter, missing a letter, adding a letter, and swapping two letters. By varying the number of LSH functions, we evaluate the uni-gram search accuracy under these four types of spelling errors. We define this precision using the ratio $P = M'/M$, where $M'$ represents the number of keywords that fall into the same hash bucket under both erroneous and correct conditions, and $M$ denotes the total number of keywords. The results are presented in Table~\ref{table3}, which illustrates the complex interrelationships among the number of LSH functions, error type, and search accuracy.

\textbf{2) Accuracy assessment of fuzzy search:} We evaluate the effectiveness of the fuzzy search accuracy assessment scheme. We define accuracy using the ratio $P = M'/M$, where $M'$ represents the number of keywords that fall into the same hash bucket under both incorrect and correct conditions, and $M$ represents the total number of keywords. We first verify the retrieval accuracy under different keywords. VeriFuzzy employs uni-gram+LSH for fuzzy processing, thereby avoiding the false positives inherent in LSH (the fundamental cause of the low accuracy of EDMF\cite{Zhong2020Efficient}), and combines this with VBTree path intersection retrieval to improve retrieval accuracy for multiple keywords (the DVFKF\cite{cuidynamic2025} scheme can achieve similar accuracy using Merkle trees). We defined four error types to validate retrieval accuracy under special scenarios. As shown in the table \ref{table3}, VeriFuzzy and DVFKF\cite{cuidynamic2025} have the same accuracy rate, while EDMF\cite{Zhong2020Efficient} lacks accuracy due to its use of bi-grams for vectorization. However, overall, our VeriFuzzy achieves high accuracy in fuzzy multi-keyword retrieval.

\begin{table}[]
\caption{Accuracy comparison of fuzzy multi-keyword search. T1: Letter substitution (e.g.,``secare” →``secure”) ; T2: Letter omission/addition (e.g.,``secre” → “secure”) ; T3: Reversed order (e.g., ``secrue” → “secure”); T4: Same root (e.g., ``walking” → “walk”).}
\label{table3}
\centering
    \resizebox{0.5\textwidth}{!}{
\begin{tabular}{cccc|cccc}
\hline
    Schemes & \multicolumn{3}{c|}{Multi-keyword accuracy rate} & \multicolumn{4}{c}{Four types of mistakes}\\ 
    ~ & kw=1 & kw=2 & kw=5 & T1 & T2 & T3 & T4  \\ \hline
    VeriFuzzy & 91\% & 87\% & 78\% & 92\% & 88\% & 100\% & 90\%  \\ 
        DVFKF\cite{cuidynamic2025} & 92\% & 85\% & 72\% & 92\% & 88\% & 100\% & 90\%  \\ 
        EDMF\cite{Zhong2020Efficient} & 85\% & 75\% & 65\% & 85\% & 80\% & 70\% & -  \\ \hline

\end{tabular}
}
\end{table}

\textbf{3) Update:} In the \textbf{IndexGen} phase, VeriFuzzy is similar to the comparison scheme, calculating the insertion values of document index keywords through $k$ hash functions and pseudo-random functions. In the \textbf{Update} phase, since a logically full binary tree is maintained, the file identifier generated during the update is a binary path, and the complexity of any node is $O(\log n)$, with corresponding deletion consistency. In contrast, EDMF\cite{Zhong2020Efficient} computes the product of vector matrices while maintaining the tree, which is more time-consuming compared to our approach, which only performs hash computations; DVFKF\cite{cuidynamic2025} additionally maintains node partition indexes, resulting in higher performance overhead. The experimental results are shown in the Fig.~\ref{Evaluation} (a, b, c), particularly when the number of documents reaches 3,500, the insertion time for VeriFuzzy is only 47 ms.

\subsection{Search Performance \textbf{(RQ2)}}\label{RQ2}

The search process involves trapdoor generation and index traversal.
During trapdoor generation, VeriFuzzy and PCKS-FS\cite{Liutowards2025} perform only $q$ hash operations and pseudo-random function evaluations, while PCQS\cite{li2017adaptively} requires an additional $k$ hash operations.
When traversing the index, PCKS-FS\cite{Liutowards2025}, PCQS\cite{li2017adaptively}, and VFSA\cite{tong2023verifiable} require traversing the tree, and each node must verify the existence of the keyword. Our VeriFuzzy requires $k$ hashes per keyword for obfuscation when querying $q$ keywords, followed by hash index localization via historical versions and path verification via VBTree. Thus, our complexity is consistent with the aforementioned comparison schemes, but due to the pruning strategy, VeriFuzzy faces the worst-case scenario with a lower probability. In contrast, DVFKF \cite{cuidynamic2025} has linear complexity due to its chained retrieval. The experimental results are illustrated in Fig.~\ref{Evaluation} (d,e). Our VeriFuzzy shows advantages in search time when faced with varying numbers of query keywords. As the number of documents increases, our VeriFuzzy exhibits a sublinear growth trend in search time and achieves an improvement in search efficiency of approximately 41\% when the document count $n = 500$.

\subsection{Verification Overhead \textbf{(RQ3)}}\label{RQ3}
In terms of verification efficiency, MVSSE\cite{liu2020multi} and MVDSE\cite{xuforward2025} require the calculation of two accumulators during the verification process, with a total time complexity of $O(R(w)) + O(n)$, where $R(w)$ represents the number of documents containing the keyword $w$. Given that $R(w)$ is much smaller than the total number of documents $n$. DVFKF\cite{cuidynamic2025} and VFSA\cite{tong2023verifiable} require verification of document integrity and index integrity. During the index integrity phase, hash operations must be performed on the indexes of $n$ documents to generate verification values. In contrast, VeriFuzzy reconstructs the search binary tree based on verification information and recalculates the ciphertext summary information to determine the consistency of the returned results, which is only $O(\log n)$.
The experimental results in Fig.~\ref{Evaluation}(f) show that the retrieval verification process for 500 documents takes less than 30 milliseconds, which is significantly faster than existing schemes.

\subsection{Blockchain Performance Overhead}
Since the blockchain is primarily used as an immutable storage database in this paper, the performance analysis focuses on testing three contract components: trapdoor generation, search, and verification. Based on Caliper V0.4.0\footnote{https://github.com/hyperledger/caliper.git}, the relevant contracts were tested 1,000 times, and the results are shown in the table \ref{table4}. The results indicate that the search operation requires more time but remains within an acceptable range, while trapdoor generation and verification consume fewer resources.

\begin{table}[]
\caption{Smart Contract Operation Resource Consumption}
\label{table4}
\begin{center}
\setlength{\tabcolsep}{2.7mm}{
\begin{tabular}{cccc}
\hline

Smart Contract   &    Time (ns)   & CPU (\%)  &  Storage (MB) \\ \hline
Trapdoor Generation   & 1.94  &  1.77  &  2.33 \\
Search & 7.65 &   3.44 &   12.45 \\
Verification  &  3.27   & 1.61   & 2.35 \\ \hline
        
\end{tabular}
}
\end{center}
\end{table}

The experimental results provide conclusive answers to the aforementioned research questions, summarized as follows:
\begin{itemize}
    \item \textbf{Answer 1. }Our proposed scheme effectively handles four common types of spelling errors with a maximum search accuracy of over 80\%. Furthermore, the scheme supports dynamic file addition with a time complexity of $O(\log n)$.
    \item \textbf{Answer 2.} The search functionality of our scheme achieves a time complexity of $O(\log n)$, and compared to existing solutions, the search efficiency is improved by approximately 41\%.
    \item \textbf{Answer 3.} The verification algorithm in our solution demonstrates superior performance in terms of time overhead, with a time complexity of $O(\log n)$. This represents a significant improvement over the $O(n)$ complexity of baseline solutions.
\end{itemize}


\section{Discussion and Future work}\label{discussion}


Although VeriFuzzy demonstrates significant advantages in integrating fuzzy search, dynamic updates, and verifiable results, we acknowledge several limitations that indicate directions for future research.

\subsection{Limitations and Practical Considerations}

\textbf{Multi-User Support:} The current VeriFuzzy implementation focuses on single-user scenarios, which aligns with many practical enterprise deployment models where a single data owner manages the encrypted data. However, extending VeriFuzzy to multi-user settings would require additional mechanisms for key management and access control. Future work could explore attribute-based encryption or proxy re-encryption techniques to enable secure multi-user access while maintaining our core efficiency guarantees.

\textbf{Language-Specific Optimizations:} Our evaluation primarily uses English textual datasets, leveraging unigram-based LSH for fuzzy matching. Although the fundamental techniques are language-agnostic, specific languages (particularly character-based languages like Chinese) may require specialized tokenization and similarity metrics. The modular design of VeriFuzzy's fuzzy processing component makes it amenable to integrating language-specific preprocessing pipelines.

\textbf{Blockchain Performance Considerations:} Although our blockchain integration achieves $O(\log n)$ verification complexity, real-world deployment requires careful consideration of transaction costs and latency. The permissioned blockchain approach used in VeriFuzzy (Hyperledger Fabric) provides better performance than public blockchains, but organizations must still evaluate the trade-offs between verification assurance and operational costs. Future work could explore layer-2 solutions or optimized consensus mechanisms for better scalability.

\subsection{Broader Implications}

The VeriFuzzy framework represents a significant step toward practical encrypted search systems that do not force compromises between functionality, efficiency, and security. Our work demonstrates that through careful architectural design, it is possible to achieve strong security guarantees without sacrificing performance—a critical requirement for real-world cloud adoption.

From a systems perspective, VeriFuzzy shows how blockchain technology can be effectively integrated into searchable encryption not as a panacea, but as a targeted solution for specific trust challenges. This principled approach to technology integration provides a template for future work combining cryptographic techniques with distributed systems.

\section{Conclusion}\label{sec:conclusion}


This paper presents VeriFuzzy, a dynamic verifiable fuzzy search service framework that overcomes fundamental barriers in encrypted search integration. By introducing semantic-decoupled indexing, VeriFuzzy enables efficient $O(\log n)$ fuzzy search on Virtual Binary Trees while eliminating the token explosion problem. Our blockchain-reconstructed verification achieves $O(\log n)$ completeness checks through smart contract-based path regeneration, and leakage-resilient protocols preserve privacy in fuzzy settings. Evaluations demonstrate practical performance: 47ms updates, sub-30ms verification, and 41\% faster search than state-of-the-art approaches while maintaining IND-CKA2 security. VeriFuzzy establishes that functionality, efficiency, and security need not be mutually exclusive in encrypted search systems.

\textbf{Artifact Availability:} Our implementation is publicly available at \url{https://github.com/JackAugust/VeriFuzzy.git} to support reproducibility and future research.

%
%
%
%

\bibliographystyle{IEEEtran}
\bibliography{reference}






\end{document}